\documentclass[10pt,a4paper,conference]{IEEEtran}
\usepackage{times}
\usepackage[dvips]{graphicx}
\usepackage{latexsym}
\usepackage{mathrsfs}
\usepackage{amssymb}
\usepackage{amsmath}
\usepackage{verbatim}
\usepackage{flushend}
\allowdisplaybreaks
\usepackage{color}

\usepackage[bookmarks=false]{hyperref}
\usepackage{multirow}
\renewcommand{\QED}{\QEDopen}
\IEEEoverridecommandlockouts
\renewcommand{\baselinestretch}{1}%0.99
\begin{document}
%\breakauthorline{}% breaks lines after the n-th author
\title{FDD Massive MIMO Uplink and Downlink Channel Reciprocity Properties: Full or Partial Reciprocity?}

\author{
\authorblockN{Zhimeng Zhong$^*$, Li Fan$^*$, Shibin Ge$^*$\\}
\authorblockA{
$^*$ Wireless Network RAN Research Department, Shanghai Huawei Technologies CO., Ltd, Shanghai, China, \\
Email: \{zhongzhimeng, fanli4, geshibin\}@huawei.com\\
}
\vspace{-8mm}
}

% make the title area
\maketitle \vspace{-12mm}
\begin{abstract}
One challenge for FDD massive MIMO communication system is how to obtain the downlink channel state information (CSI) at the base station. Except for traditional codebook feedback through uplink pilot transmission, some channel reciprocity properties can be utilized through uplink channel estimation and channel parameter estimation algorithms.  In this paper, the uplink and downlink channel reciprocity properties are analyzed. It is theoretically proved that not all multipath parameters for FDD downlink and uplink channels are equivalent. Therefore, the so called full reciprocity property does not hold while the partial reciprocity property holds. Moreover, the channel measurement campaign is conducted to verify our theoretical analysis. %Finally, in order to support the partial reciprocity property, the revision for the standardization 5G channel model is proposed as well. 
With the contribution of this paper, the FDD massive MIMO system transmission scheme design could be led to the right direction.
\end{abstract}


%\IEEEpeerreviewmaketitle

\section{Introduction}

Massive multiple-input multiple-output (MIMO) will be an essential part of 5G systems. Knowledge of channel state information (CSI) at the transmitter (CSIT) is a fundamental prerequisite for operation of massive MIMO systems. However, a massive MIMO base station (BS) has much larger number of antennas than users'. In time division duplex (TDD) systems, the BS can obtain the downlink CSIT through uplink pilot transmission from the user equipment (UE), since the channel reciprocity holds as long as uplink and downlink transmissions occur within the channel coherence time. However, in frequency division duplex (FDD) system, the uplink (UL) and downlink (DL) channels have no such reciprocity as TDD system, since the uplink  and downlink  in FDD system are usually separated by more than a coherence frequency bandwidth. The current 5G NR system uses downlink CSI-RS transmission, and Type I or Type II codebook feedback from the UE, to get downlink CSIT \cite{c1}. It leads to considerable feedback overhead, and performance loss due to quantized error and channel aging problem \cite{c2}. Recently, one alternative method to get CSIT is to use some FDD UL \& DL channel intrinsic reciprocity properties. There are two kinds of assumption on such FDD channel reciprocity. One assumes that the channel consists of numerous multipath, and all the parameters of each multipath, (including phase, amplitude, delay, angle of arrival, departure, etc.) are equivalent to UL and DL channels. Such assumption is referred to as full reciprocity property. Based on the full reciprocity property, one method to obtain the DL CSIT is the extrapolation of the complex, instantaneous channel frequency response by UL CSI without any feedback in \cite{c3,c4,c5,c6,c7}. However, In\cite{c8,c9}, some measurement results and theoretical investigations show that the full reciprocity does not hold, and it points out that the phase relationship between different multipath components is not reciprocal for FDD DL and UL channels. Therefore, another assumption assumes that only part of channel characteristics have such reciprocity property, such as, the angular power spectrum, channel covariance matrix, delay and angle of each multipath, etc, which is called partial reciprocity property. Some literatures make use of such partial reciprocity property with limited feedback to improve DL massive MIMO transmission performance \cite{c10,c11,c12,c13,c14,c15,c16}.

In this paper, from the point of view of wireless propagation aspect, multipath parameters' variations with frequency are theoretically analyzed, and it is proved that the full reciprocity property does not hold through our theoretical analysis, but the FDD UL and DL channels have some partial reciprocity properties. Different from \cite{c8,c9}, intensively theoretical analysis based on electromagnetic theory is provided in this paper. Moreover, some channel measurement campaigns are conducted, and the partial channel reciprocity is proved through the realistic channel impulse response. 
%Finally, the standardization 5G channel model in \cite{c17} is analyzed, and it is found that it needs minor revisions to support the FDD channel partial reciprocity properties.

Notations: Vector and matrices are denoted by bold lowercase and uppercase letters, respectively. Superscripts $(\cdot)^*$, $(\cdot)^T$ and $(\cdot)^H$ stand for conjugate, transpose and Hermitian transpose operators.
\begin{figure}[th]
\centerline{\includegraphics[width=8.8cm,height=8.8cm]{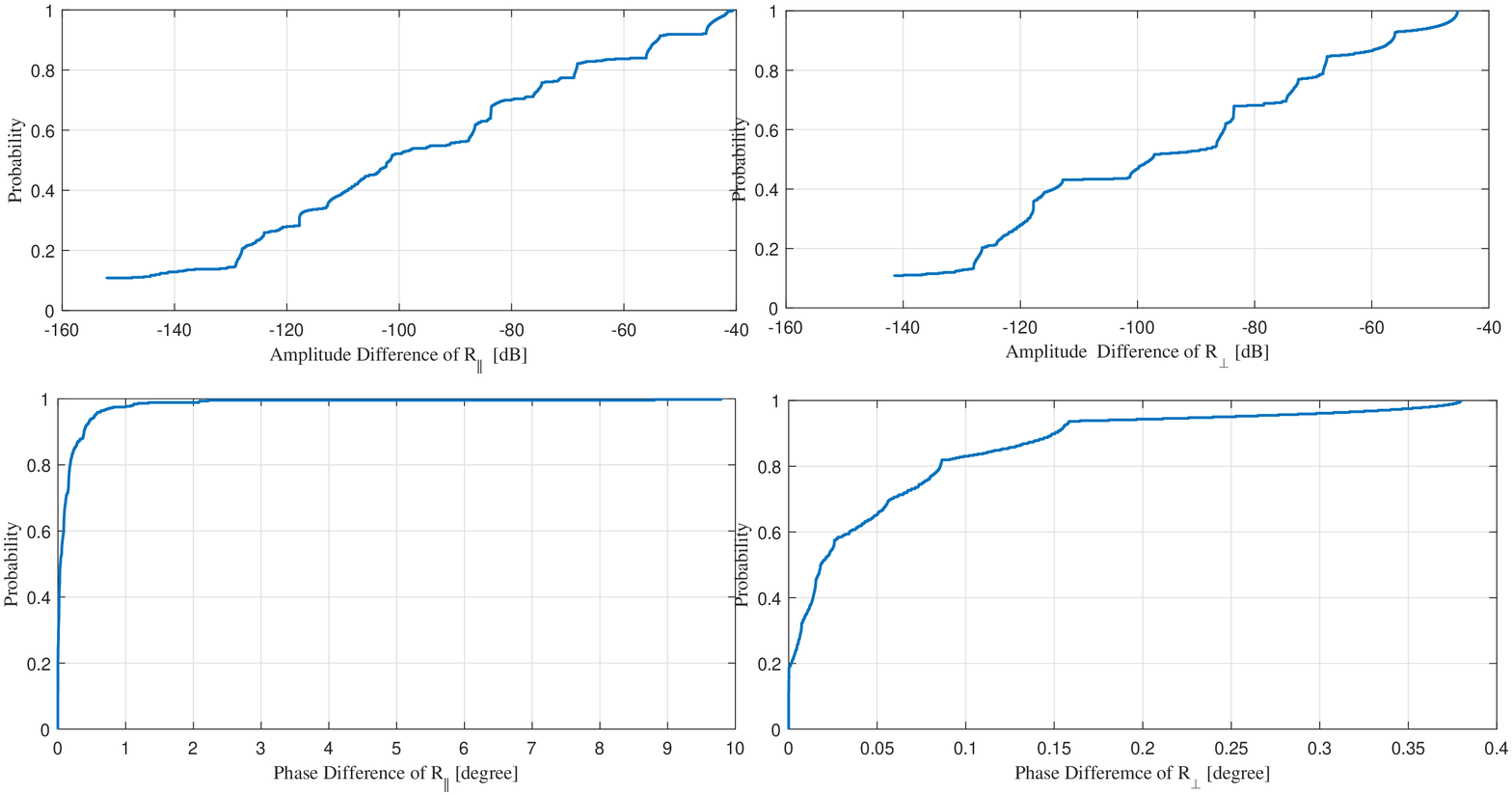}}
\caption{. The CDF of ampitude and phase difference for different materials' diffraction coefficients at 1.8 GHz and 1.99 GHz.}
\label{fig1}
\end{figure}

\section{Channel Model and Theoretical Analysis on FDD Reciprocity}
\subsection{Channel Model Description}

Considering the dual-polarization narrow band MIMO channel matrix with the dimension of $2M \times 2N$, which can be modeled as:
\begin{equation}
\mathbf{H}(t)=\left[\begin{array}{ll}{\mathbf{H}^{(1,1)}(t)} & {\mathbf{H}^{(1,2)}(t)} \\ {\mathbf{H}^{(2,1)}(t)} & {\mathbf{H}^{(2,2)}(t)}\end{array}\right],
\label{eq1}
\end{equation}
where,\\$\mathbf{H}^{(p, q)}(t)=\sum\limits_{l=1}^{L} \alpha_{l}^{(p, q)} \exp (-j 2 \pi f \tau_{l}) \exp (j 2 \pi \lambda^{-1} v_{l} t) \mathbf{a}_{r x, l} \mathbf{a}_{t x, l}^{T}$
$p, q \in\{1,2\}$ are the types of receiver and transmitter antenna elements' polarization, respectively; $L$ is the number of multipath; $\alpha_{l}^{(p, q)}$ is the fading coefficient of the $l$th multipath when the transmission antenna and receiver antenna polarizations' are $q$ and $p$;  $f$ is the subcarrier frequency; $\tau_l$ is the delay of the $l$th multipath; $\lambda$ is the wavelength; $v_l$ is the movement speed of the $l$th multipath; $\mathbf{a}_{r x, l}\left(\theta_{l}, \varphi_{l}\right)$ and $\mathbf{a}_{t x, l}\left(\vartheta_{l}, \mu_{l}\right)$ are the antenna steer vectors for AOA (Azimuth angle Of Arrival) $\varphi_{l}$, EOA (Elevation angle Of Arrival) $\theta_{l}$ at receiver, and AOD (Azimuth angle of Departure) $\mu_l$, EOD (Elevation angle of Departure) $\vartheta_{l}$ at transmitter, the dimensions of which are $M \times 1$ and $N \times 1$, respectively. It should be noted that the specific form of $\mathbf{a}_{r x, l}\left(\theta_{l}, \varphi_{l}\right)$ and $\mathbf{a}_{t x, l}\left(\vartheta_{l}, \mu_{l}\right)$ are determined by the antenna array geometry. One can refer to the Eq. (15) in \cite{c18} for the formulations of $\mathbf{a}_{r x, l}\left(\theta_{l}, \varphi_{l}\right)$ and $\mathbf{a}_{t x, l}\left(\vartheta_{l}, \mu_{l}\right)$ in details.
\subsection{Theoretical Analysis on FDD Reciprocity Properties}
From \eqref{eq1}, it can be seen that the MIMO channel matrix is made up of multipath channel parameters which include $\alpha_{l}^{(p, q)}$, $\tau_l$, $v_l$, $\left(\theta_{l}, \varphi_{l}\right)$ and $\left(\vartheta_{l}, \mu_{l}\right)$. In the following, theoretical analysis shows that only part of these channel parameters have FDD reciprocity property.
\par \emph{Remark 1: Based on geometrical optics theory, it is straightforward to have a conclusion that one arbitrary channel path's angles at BS$\alpha_{l}^{(p, q)}$, $\tau_l$, $v_l$, $\left(\theta_{l}, \varphi_{l}\right)$ and UE$\left(\vartheta_{l}, \mu_{l}\right)$ must be reciprocal for FDD DL and UL channels. Additionally, $v_l$ is only determined by the scatter's movement so that it is equivalent for FDD DL and UL channels as well.}
\begin{figure}[h]
\centerline{\includegraphics[width=9cm,height=9.6cm]{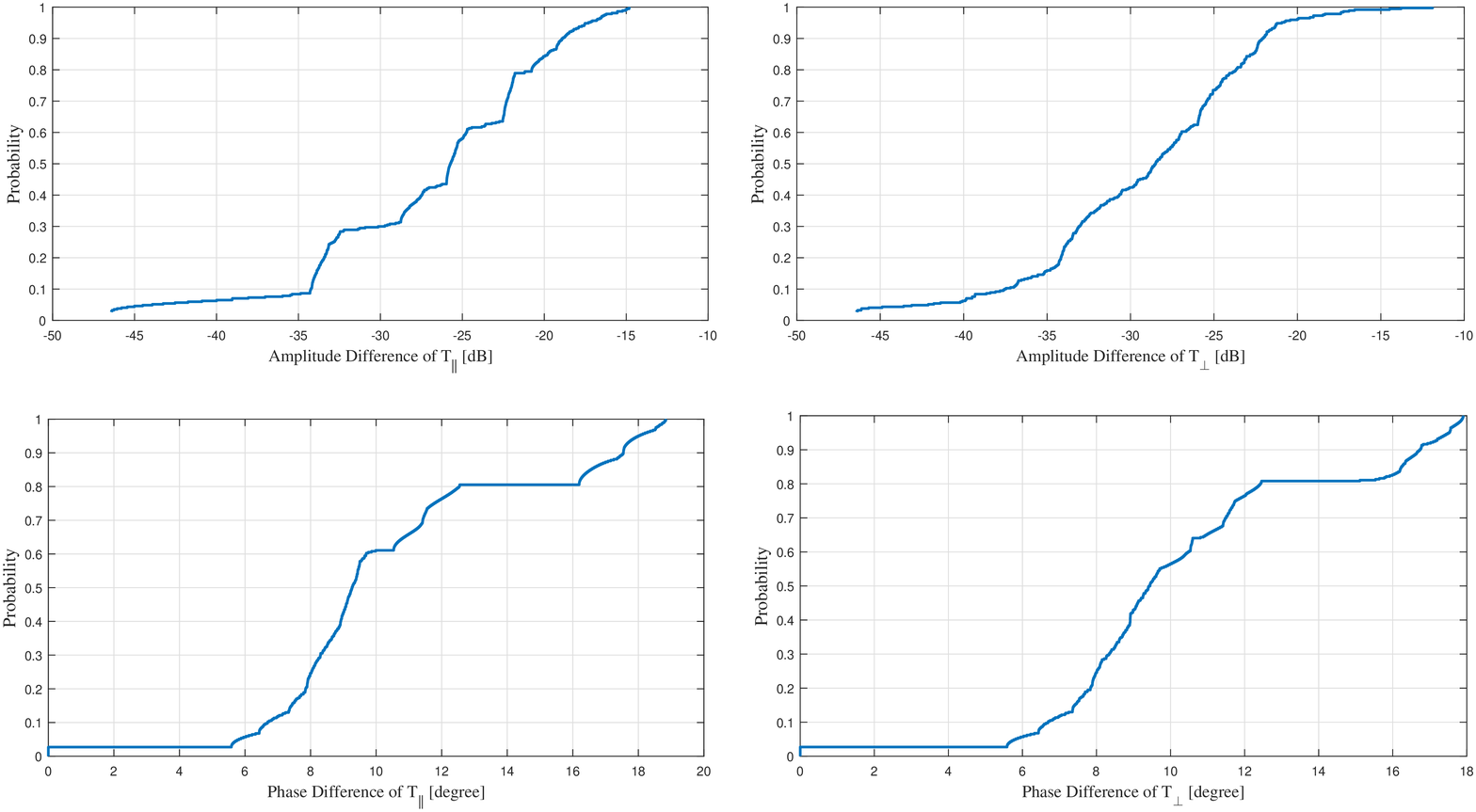}}
\caption{. The CDF of ampitude and phase difference for different materials' diffraction coefficients at 1.8 GHz and 1.99 GHz.}
\label{fig2}
\end{figure}
\par Secondly, the delay for each multipath $\tau_l$ is determined by the distance for each path and the speed of light, which can be written as
\begin{equation}
\tau_{l}=\frac{d_{l}}{v}
\label{eq2}
\end{equation}
where, $d_l$ is the distance from the transmitter to receiver for multipath $l$; $v$ is the speed of light in medium, which is equal to $v=C/n$, where $C$ is the speed of light in vacuum, and $n \approx \sqrt{\varepsilon_{r} \mu_{r}}$ is refractive index in different medium which is determined by relative permittivity $\varepsilon_{r}$ and relative permeability $\mu_r$. According to \cite{c19}, $\varepsilon_{r}$ and $\mu_r$ are frequency independent within FDD duplex frequency range, so we have the following remark:
\par \emph{Remark 2: The delay $\tau_l$ for each multipath has the FDD reciprocity property.}
\par Next, we will elaborate the frequency-dependent property of $\alpha_{l}^{(p, q)}$ in details. $\alpha_{l}^{(p, q)}$ could be modeled as
\begin{equation}
\alpha_{l}^{(p, q)}=\left[\begin{array}{c}{F_{r, \theta}^{(p)}\left(\theta_{l}, \varphi_{l}\right)} \\ {F_{r, \varphi}^{(p)}\left(\theta_{l}, \varphi_{l}\right)}\end{array}\right]^{T} \mathbf{A}_{l}\left[\begin{array}{c}{F_{t, \theta}^{(q)}\left(\vartheta_{l}, \mu_{l}\right)} \\ {F_{t, \varphi}^{(q)}\left(\vartheta_{l}, \mu_{l}\right)}\end{array}\right]
\label{eq3}
\end{equation}
where, $F_{r, \theta}^{(p)}$ and $F_{r, \varphi}^{(p)}$ are the field patterns of receive antenna with polarization type $p$, and in the direction of the spherical basis vectors, $\theta$ and $\varphi$, respectively; $F_{t, \theta}^{(q)}$ and $F_{t, \theta}^{(q)}$ are the antenna patterns of transmit antenna with polarization type $q$, and in the direction of the spherical basis vectors, $\theta$ and $\varphi$ respectively; $\mathbf{A}_{l}$ is the $l$th multipath $2 \times 2$ depolarization matrix, which can be written by \cite{c17}
\begin{equation}
\mathbf{A}_{l}=\left[\begin{array}{cc}{e^{j \emptyset_{l}^{\theta \theta}}} & {\sqrt{\kappa_{l}^{-1}} e^{j \emptyset_{l}^{\theta \phi}}} \\ {\sqrt{\kappa_{l}^{-1}} e^{j \emptyset_{l}^{\phi \theta}}} & {e^{j \emptyset_{l}^{\phi \phi}}}\end{array}\right]
\label{eq4}
\end{equation}
where, $\kappa_{l}$ is the cross polarization power ratios (XPR) for each path $l$, and $\left\{\emptyset_{l}^{\theta \theta}, \emptyset_{l}^{\theta \phi}, \emptyset_{l}^{\phi \theta}, \emptyset_{l}^{\phi \phi}\right\}$ are the phases, which are caused by reflection, diffraction, and transmission depolarization effects.
\begin{figure}[t]
\centerline{\includegraphics[width=9.7cm]{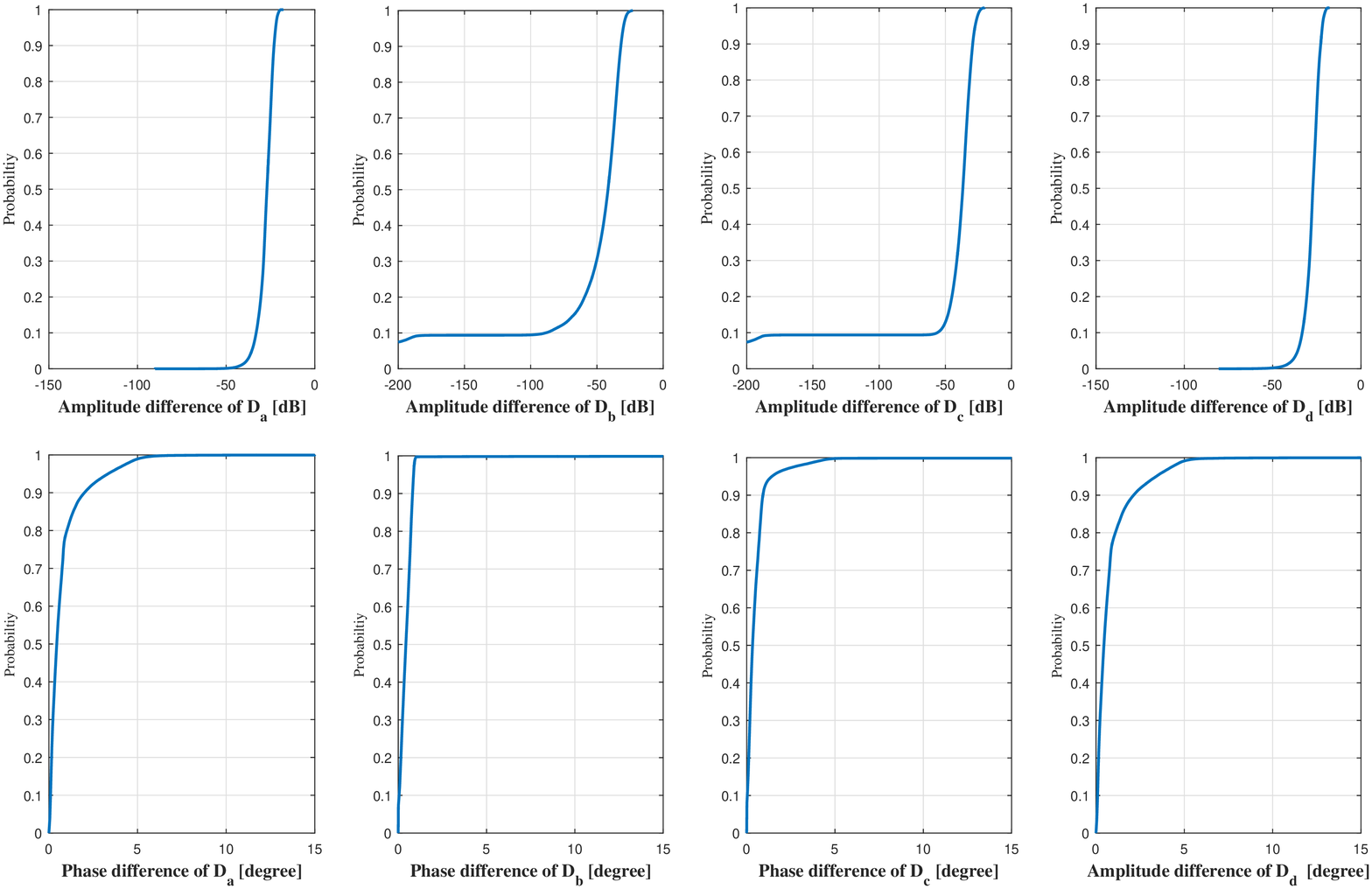}}
\caption{. The CDF of ampitude and phase difference for different materials' diffraction coefficients at 1.8 GHz and 1.99 GHz.}
\label{fig3}
\end{figure}
\par According to \cite{c20}, the depolarization matrix $\mathbf{A}_{l}$ for reflection, diffraction, transmission can be modeled as
\begin{equation}
\mathbf{A}_{l}=\left[\begin{array}{ll}{\alpha_{\theta \theta}} & {\alpha_{\theta \phi}} \\ {\alpha_{\phi \theta}} & {\alpha_{\phi \phi}}\end{array}\right]
\label{eq5}
\end{equation}
\par The entries of matrix $\mathbf{A}_{l}$ for reflection can be calculated as Eqs. (6-17) in \cite{c20}, and one can see that these entries of $\mathbf{A}_{l}$ are only determined by $R_{\|}$ and $R_{\perp}$, the Fresnel reflection coefficients for parallel polarization and perpendicular polarization, respectively. Similar to the reflection , the entries of matrix $\mathbf{A}_{l}$ for transmission are determined by $T_{\|}$ and $T_{\perp}$ \cite{c21}, the Fresnel transmission coefficients for parallel polarization and perpendicular polarization, respectively. Therefore, it can be seen that FDD reciprocity properties of the depolarization matrix for reflection and transmission are only determined by $R_{\|}$, $R_{\perp}$, $T_{\|}$ and $T_{\perp}$. According to \cite{c20,c21}, $R_{\|}$, $R_{\perp}$, $T_{\|}$ and $T_{\perp}$ can be respectively written as
\begin{equation}
R_{\|}=\frac{-\sqrt{\frac{\mu_{1}}{\varepsilon_{1}}} \cos \theta_{i}+\sqrt{\frac{\mu_{2}}{\varepsilon_{2}}} \cos \theta_{t}}{\sqrt{\frac{\mu_{1}}{\varepsilon_{1}}} \cos \theta_{i}+\sqrt{\frac{\mu_{2}}{\varepsilon_{2}}} \cos \theta_{t}}
\label{eq6}
\end{equation}
\begin{equation}
R_{\perp}=\frac{\sqrt{\frac{\mu_{2}}{\varepsilon_{2}}} \cos \theta_{i}-\sqrt{\frac{\mu_{1}}{\varepsilon_{1}}} \cos \theta_{t}}{\sqrt{\frac{\mu_{2}}{\varepsilon_{2}}} \cos \theta_{i}+\sqrt{\frac{\mu_{1}}{\varepsilon_{1}}} \cos \theta_{t}}
\label{eq7}
\end{equation}
\begin{equation}
T_{\|}=\frac{2 \sqrt{\frac{\mu_{2}}{\varepsilon_{2}}} \cos \theta_{i}}{\sqrt{\frac{\mu_{1}}{\varepsilon_{1}}} \cos \theta_{i}+\sqrt{\frac{\mu_{2}}{\varepsilon_{2}}} \cos \theta_{t}}
\label{eq8}
\end{equation}
\begin{equation}
T_{\perp}=\frac{2 \sqrt{\frac{\mu_{2}}{\varepsilon_{2}}} \cos \theta_{i}}{\sqrt{\frac{\mu_{2}}{\varepsilon_{2}}} \cos \theta_{i}+\sqrt{\frac{\mu_{1}}{\varepsilon_{1}}} \cos \theta_{t}}
\label{eq9}
\end{equation}
where, assuming the wave travels to the planar interface formed by two lossless media, and $\varepsilon_{1}$, $\mu_{1}$, $\varepsilon_{2}$, $\mu_{2}$ are the two lossless media's permittivity, and permeability, respectively; $\theta_i$ and $\theta_t$ are the incident and transmission angles. Considering the wave penetrates the media so that the total transmission coefficients should be
\begin{equation}
T_{\|}^{t o t a l}=T_{\|}^{1} T_{\|}^{2}
\label{eq10}
\end{equation}
\begin{equation}
T_{\perp}^{t o t a l}=T_{\perp}^{1} T_{\perp}^{2}
\label{eq11}
\end{equation}
where $T_{\|}^1$, $T_{\perp}^1$ are equal to $T_{\|}$ and $T_{\perp}^1$ in \eqref{eq8}\eqref{eq9}; $T_{\|}^2$, $T_{\perp}^2$ are similar to $T_{\|}^1$, $T_{\perp}^1$, but the transmission angle and material properties are different, of which the details can be referred to \cite{c21} due to the limitation of space.
\begin{figure}[tb]
\centerline{\includegraphics[width=9cm]{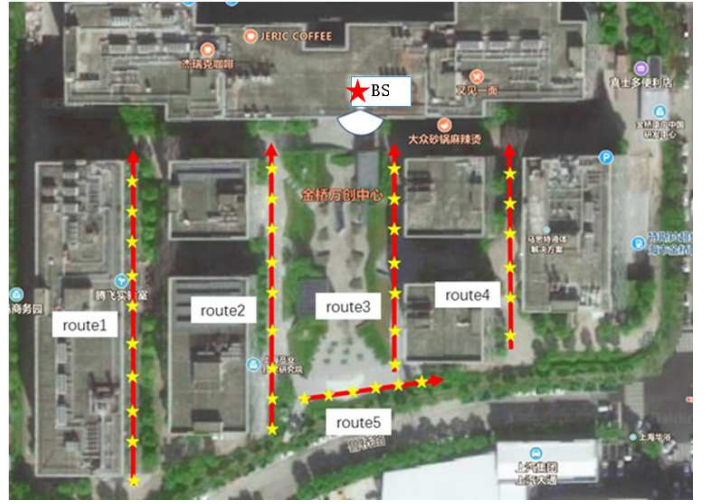}}
\caption{. The birdview of channel measurement campaign scenarios.}
\label{fig4}
\end{figure}

\par Herewith, it is assumed that UL and DL carrier frequencies are 1.8G Hz and 1.99 GHz, respectively. Based on the material properties in Table 3, \cite{c19} and equations \eqref{eq6}--\eqref{eq9}, the reflection and transmission coefficients for all the  materials are calculated at 1.8G Hz and 1.99 GHz. Then the CDFs (Cumulative Distribution Function) of the relative differences of $R_{\|}$, $R_{\perp}$ are given in Fig. \ref{fig1}. One can see that the phase and amplitude relative differences are very close to zero from Fig. \ref{fig1}. Therefore, we have the following remark:
\par \emph{Remark 3: Since the reflection coefficients are frequency-independent, the depolarization matrix for reflection has FDD reciprocity property.}

\par Similar to reflection, the total transmission coefficients can be also calculated and compared with different frequency. The CDFs of the relative difference of $T_{\|}^{t o t  a l}$, $T_{\perp}^{t o t a l}$ are given in Fig. \ref{fig2}. It can be seen that the amplitude relative difference is very close to zero, but the CDF of relative phase difference shows that the max phase difference could be larger than 15 degree. Hence, we have the conclusion as follows:
\par \emph{Remark 4: Since the transmission coefficients are frequency-dependent, the depolarization matrix for transmission has no FDD reciprocity property.}
\par Except for reflection and transmission, another important propagation mechanism is diffraction. The depolarization matrix for diffraction can be written as \cite{c20}
\begin{equation}
\mathbf{A}_{l}=\left[\begin{array}{ll}{D_{a}} & {D_{b}} \\ {D_{c}} & {D_{d}}\end{array}\right]
\label{eq12}
\end{equation}
where one can refer to the detailed expressions of $D_a$, $D_b$, $D_c$, and $D_d$ in Eqs. (6--31) \cite{c20}. In Fig. \ref{fig3}, it shows the CDFs of the relative differences of $D_a$, $D_b$, $D_c$, $D_d$ at 1.8G Hz and 1.99 GHz. Similar to Fig. \ref{fig3}, it can be seen that the amplitude relative difference is very small but the CDF of relative phase difference shows that the phase difference could be larger than 5 degree.
\par \emph{Remark 5: Since the diffraction coefficients are frequency-dependent, the depolarization matrix for diffraction has no FDD reciprocity property.}

 Each multipath is the combination of the propagation mechanisms, such as, reflection, diffraction, transmission. According to the theoretical analysis above, we can have the following conclusion:
\par Conclusion: The $2 \times 2$ depolarization matrix $\mathbf{A}_{l}$ for each multipath has no FDD reciprocity property, only if multipath is caused by specular reflection.
\begin{figure}[tb]
\centerline{\includegraphics[width=9cm,height=4.5cm]{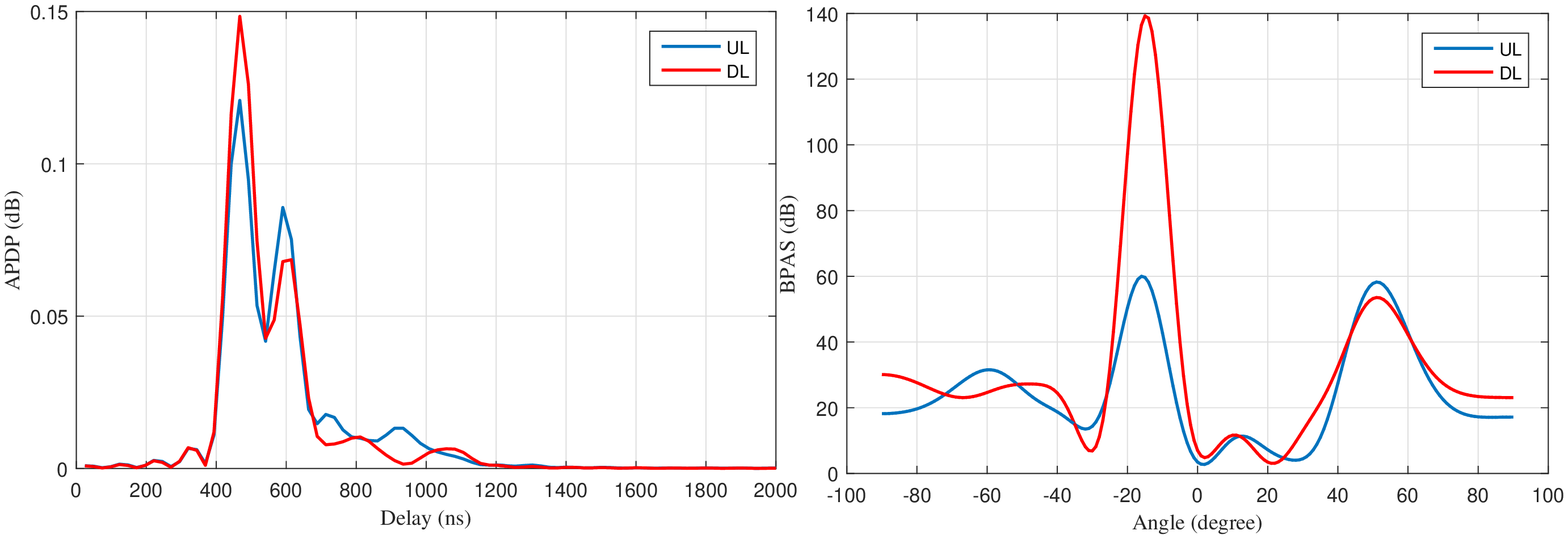}}
\caption{. The BPAS and APDP Comparison between FDD UL and DL at LOS location.}
\label{fig5}
\end{figure}
\begin{figure}[tb]
\centerline{\includegraphics[width=9cm,height=4.5cm]{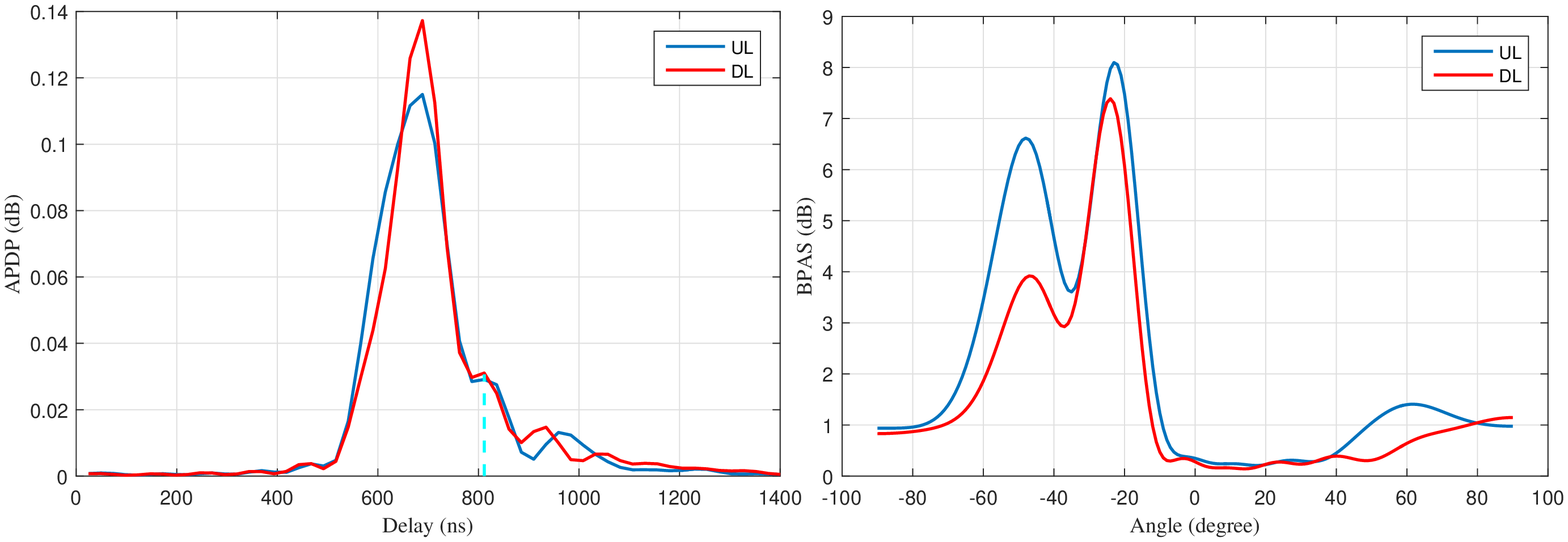}}
\caption{. The BPAS and APDP Comparison between FDD UL and DL at NLOS location.}
\label{fig6}
\end{figure}

\section{Partial Reciprocity Verification by Channel Measurement}
The channel measurement campaigns are needed to verify the theoretical analysis of the partial reciprocity in Section II. The channel sounder system was used in the measurement campaign. An MIMO uniform linear array (ULA) which has 8 patches was used at BS. Each antenna patch is separated by $d=8.33$ cm which is the half wavelength at 1.8 GHz, and has a pair of $\pm 45^{\circ}$ polarized co-located antenna elements. At UE side, one dipole antenna with vertical polarization was used. The UL and DL carrier frequencies are 1.8 GHz and 1.9 GHz, respectively. BS generates the channel sounding signals at 1.8 GHz and 1.9 GHz, and transmits these two bands' signal to UE. Both of the signal bandwidth at two carrier frequencies are 10 MHz, and a modulated PN-sequence with length 1023 was transmitted as channel sounding signal. It should be noted that such one-directional two bands transmission scheme is equivalent to FDD DL and UL transmissions, since UL and DL channels are reciprocal for TDD system \cite{c8}. Herewith, the transmission power difference for UL and DL is ignored, since we only focus on the pure propagation part.

\par The measurement campaign was conducted in Tengfei CBD, and the measurement scenario belongs to the urban macro scenario as illustrated in Fig.  \ref{fig4}. The BS height is 30 meters, which is marked by red star, and the UE is about 1.5 m height marked by yellow star as shown in Fig. \ref{fig4}. The routes L1 and L4 shown in Fig.  \ref{fig4} are NLOS cases, routes L2 and L3 are LOS cases, and L5 is LOS and NLOS hybird case. The BS is deployed  on the top of building in order to ensure the UE locations are within the BS antenna main lobe.

\par In order to verify FDD partial reciprocity, the averaged power delay profile (APDP) and the Bartlett Power Angle Spectrum (BPAS) are used as the figure of merits. The APDP and BPAS can be respectively formulated by
\begin{equation}
P_{t}(\tau)=\frac{1}{M} \sum_{m=1}^{M} \frac{|\mathbf{h}(t, \tau, m)|^{2}}{\sum_{\tau}|\mathbf{h}(t, \tau, m)|^{2}}
\label{eq13}
\end{equation}
\begin{equation}
P_{t}(\theta)=\left|\boldsymbol{a}^{H}(\theta) \mathbf{R}_{h} \boldsymbol{a}(\theta)\right|
\label{eq14}
\end{equation}
where $\mathbf{h}(t, \tau, m)$ is the channel impulse response for the $t$th TTI, $m$th BS antenna, and $M=16$ is the total antenna number at BS; $\mathbf{R}_h$ is the channel covariance matrix at BS; $\pmb{a}(\theta) = [1, e^{-j 2 \pi \frac{d}{\lambda} \cos \theta}, \ldots, e^{-j 2 \pi \frac{7 d}{\lambda} \cos \theta}, 1, e^{-j 2 \pi \frac{d}{\lambda} \cos \theta}, \ldots, $ $ e^{-j 2 \pi \frac{7 d}{\lambda} \cos \theta}]$, which is the steer vector for BS antenna and $\lambda$ is the wavelength of UL or DL.

\par
\begin{figure}[tb]
\centerline{\includegraphics[width=9cm,height=4cm]{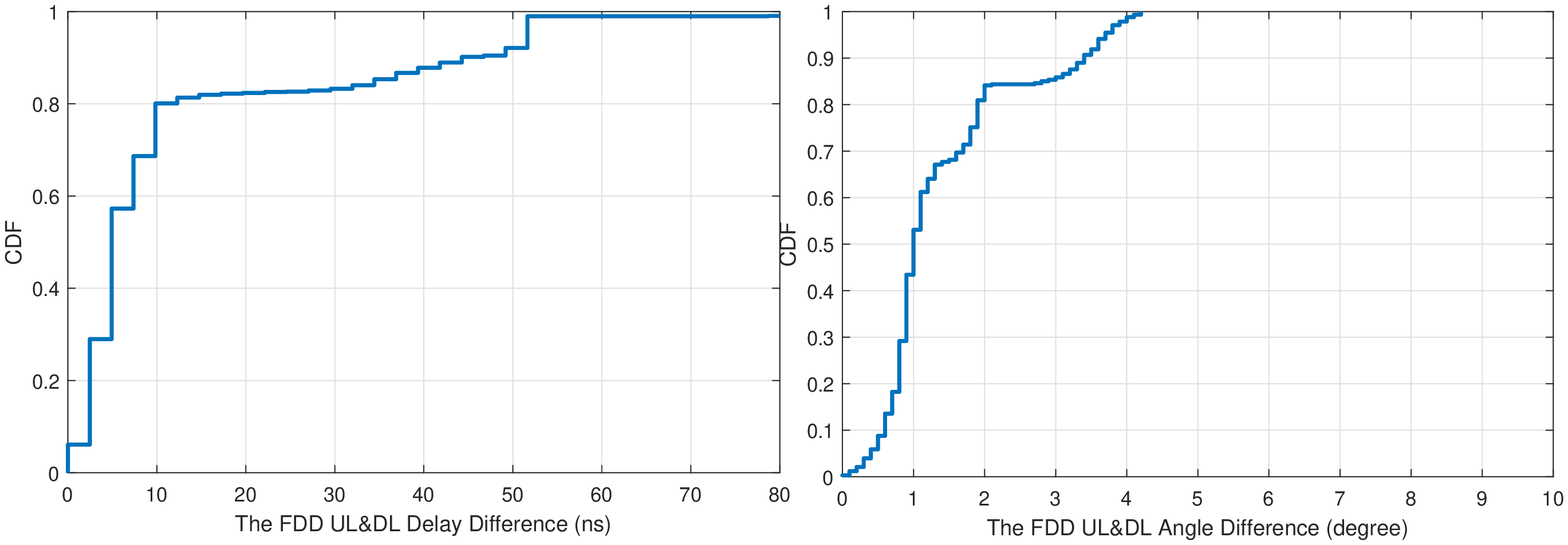}}
\caption{. The CDF of angle and delay differences for FDD UL and DL at LOS measurement case.}
\label{fig7}
\end{figure}

\begin{figure}[tb]
\centerline{\includegraphics[width=9cm,height=4cm]{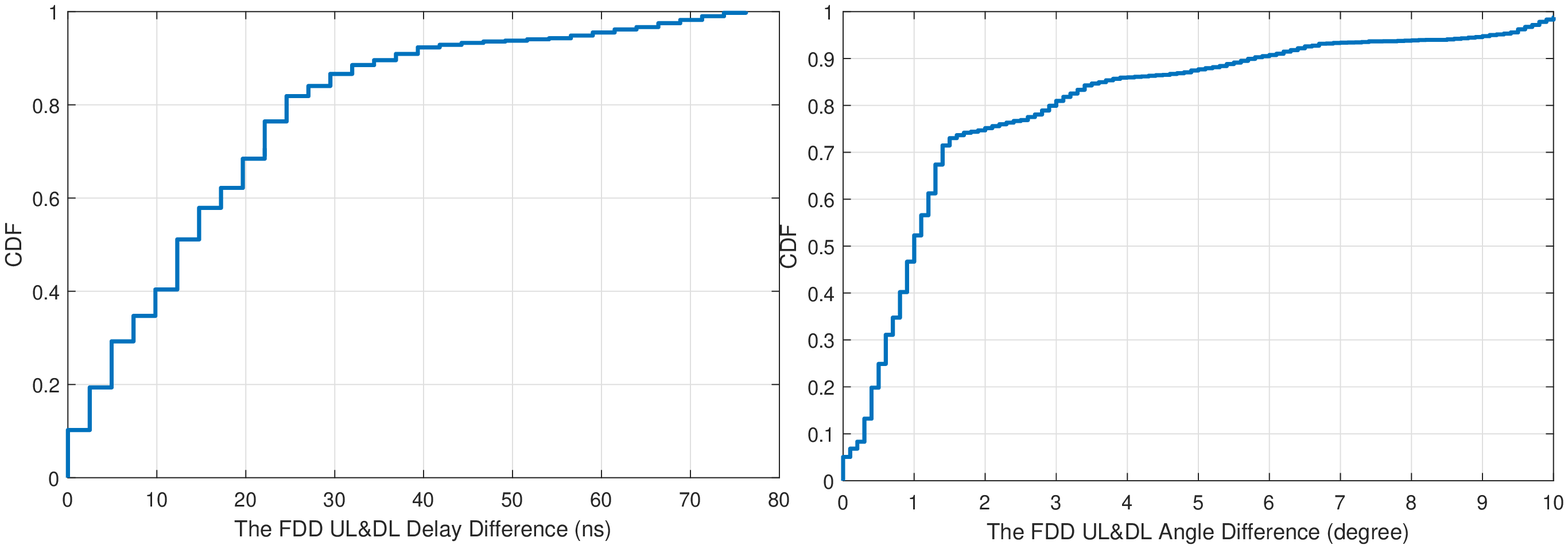}}
\caption{. The CDF of angle and delay differences for FDD UL and DL at NLOS measurement case.}
\label{fig8}
\end{figure}

 Fig. \ref{fig5} and Fig. \ref{fig6} show the BPAS and APDP of FDD UL \& DL channels at L2 route's location 5, and L1 route's location 4, respectively. From these two figures, it can be seen that the dominant paths' delay and angle for UL and DL are almost equivalent, but the amplitude may not be equivalent. This is because that the relative phase of each multipath caused by multipath delay is different due to UL and DL different carrier frequencies, as well as the depolarization matrix for each multipath is different for FDD UL \& DL in accordance with the previous analysis. In Fig. \ref{fig7} and Fig. \ref{fig8}, we give the CDF  of the FDD UL \& DL delay and angel differences for all the NLOS and LOS channel measurement data. The local maximum points in the APDP and BPAS are regarded as the multipath positions, and local maximum points in Fig. \ref{fig7} and Fig. \ref{fig8}, which of power is not less than the strongest path's relative power $-$20 dB, are selected. From Fig. \ref{fig7}, we can see that the delay difference and angle difference from FDD UL \& DL are less than 50ns and 5 degree in LOS case, respectively. For NLOS case as shown in Fig. \ref{fig8}, the delay and angle differences are less than 80 ns and 10 degree, respectively. In comparison of LOS and NLOS cases, the delay and angle differences for FDD UL \& DL in NLOS case are larger than LOS case, this is because that there are more multipath in NLOS case so that the interference between multipath would be increased, and FDD partial \ reciprocity in NLOS case becomes worse than LOS case.
%
%\section{Standarization Channel Model Revision}
%

%3GPP proposes the 5G NR channel model in \cite{c17}, and the correlation modelling for multi-frequency simulations is illustrated in Section 7.6.5 \cite{c17}. The FDD UL \& DL channel impulse response can be generated through this correlation model. However, it needs some minor revisions in order to satisfy the FDD partial reciprocity properties. In the correlation model for multi-frequency simu\emph{}lations, the angle coupling of rays within a cluster in the fast fading modelling step 8 in Section 7.6 \cite{c17}, is randomly and independent for UL and DL. In such modelling setting, one ray's angles at BS and UE sides will be different between UL channel and DL channel, which is not in line with the angle reciprocity property in Remark 1.  Therefore, it is recommended that the Step 8 would not be independently implemented for the FDD UL and DL frequency bands, in the sense that the coupling of rays for both azimuth and elevation should be equivalent for FDD DL and UL channels.
%
\section{Conclusions}
In this work, the reciprocity properties for FDD DL and UL channels are analyzed. By our theoretical analysis and some channel measurement campaign verifications, it is found that the full reciprocity property for FDD DL and UL channels dese not hold, which means that the DL channel extrapolation through UL channel without any feedback will not work in reality, even if BS can perfectly derive the channel parameters from UL channel estimation. However, it is proved that the multipath angle and delay exist reciprocity property in FDD UL and DL channels, which means the FDD partial reciprocity property holds.  This work helps to lead the FDD Massive MIMO enhancement schemes to the right direction.

\bibliographystyle{IEEEtran}
\bibliography{IEEEabrv,reference}

%\newpage

%\newpage

%\newpage

%\newpage

%\newpage

%\newpage

\end{document}